\documentclass[preprint]{JHEP} 
\usepackage{epsfig}

\relax

\def\l{\lambda}

\def\be{\begin{equation}}
\def\ee{\end{equation}}
\def\bs{\begin{subequations}}
\def\es{\end{subequations}}

\def\l{\lambda}

\def\d{\partial}

\def\sp{\;\;\;,\;\;\;}

\def\dr{\dot r}
\def\dt{\dot\varphi}

\newcommand\fverb{\setbox\pippobox=\hbox\bgroup\verb}
\newcommand\fverbdo{\egroup\medskip\noindent%
                        \fbox{\unhbox\pippobox}\ }
\newcommand\fverbit{\egroup\item[\fbox{\unhbox\pippobox}]}
\newbox\pippobox
\title{Mirage Cosmology}

\author{by A. Kehagias\\
        Physics Department, NTUA,
15773 Zografou, Athens, GREECE\\
        E-mail: \email{kehagias@mail.cern.ch}}
\author{ E. Kiritsis\\
Physics Department, University of Crete,
71003 Heraklion, GREECE\\
E-mail: \email{kiritsis@physics.uch.gr}}

\preprint{NTUA-99/74\\ \hepth{9910174}}      

\abstract{A brane universe moving in a curved higher
dimensional bulk space is considered. The motion induces a
cosmological evolution on the universe
brane that is indistinguishable from a similar one induced by
matter density on
the brane. The phenomenological implications of such an idea are discussed.
Various mirage energy densities are found, corresponding to dilute matter
driving the cosmological expansion, many having superluminal
properties $|w|>1$
or violating the positive energy condition. It is shown that energy density
due to the world-volume fields is nicely incorporated into the picture.
It is also pointed out that the
initial singularity problem is naturally resolved in this context.}

\begin{document}

\maketitle 

\section{Introduction}

In this paper we will elaborate on some ideas related to the possibility
that our observable four-dimensional world is a three-brane embedded in
ten dimensional string theory.
There has been earlier speculations \cite{misha} stating that our
observable
four-dimensional universe is a domain wall embedded in a higher
dimensional
space.
Although there is no strong theoretical or experimental
motivation for this idea, it has been
revived
recently \cite{lyk}-\cite{Ver} motivated by the possibility of large
compact internal dimensions \cite{anto} the notion and existence of
D-branes
in string theory \cite{pol}, and the fact that orientifold \cite{orie} and
D-manifold \cite{Dman} vacua of
string theory can be thought of as lower-dimensional  D-branes embedded
in a
ten-dimensional bulk.
An early example of this is the Horava-Witten picture for the
non-perturbative heterotic $E_8\times E_8$ string \cite{hw}, the
relevance
of this for gauge coupling unification upon compactification to
five dimensions \cite{wit} and the associated picture of supersymmetry
breaking \cite{hor}.

An extra motivation is the will of theorists to provide with potential
new physics signals the experimental physics community.
Here, in particular,  the new physics can be string effects, or quantum
gravitational effects at scales that are well below the four-dimensional
Planck
scale.
One can avoid typical string theory constraints by focusing on type I
(orientifold) vacua of string theory.
Although this approach has not obviously solved any major theoretical
problem yet it is an interesting alternative and its implications
should be pursued.

In the context we will assume, the Standard Model gauge bosons as well as charged
matter
arise as fluctuations of the D-branes. We can thus consider the universe
(standard model) to be living on a collection of coincident branes,
while hidden gauge interactions can be localized on other branes.
Gravity
as well as other universal interactions is living in the bulk space.

There is an approximation which is very useful in order to treat the
dynamics of the universe brane. This is the probe limit in which the
influence
of the probe brane source to the bulk fields is negligible.
This has been a natural and useful tool  \cite{prob}-\cite{2}
in order to understand issues in
the context of AdS/CFT correspondence \cite{mald}.
However, for the spherically symmetric bulk configurations we will
consider, the probe limit will give exact results.

There has been a lot of recent work on the potential cosmological models
associated to a brane universe \cite{dvty}-\cite{free}.
Our approach will be slightly different.
We can imagine the collection of other branes to provide a gravitational
background which is felt by the universe brane treated as a probe.
In this context a universe three-brane (or higher compactified brane)
can be in motion in ten-dimensional space in the
presence of a the gravitational field of the other branes.
We ignore its back reaction to the ambient
geometry.
We will show that the motion  in ambient space induces cosmological
expansion (or
contraction) on our universe simulating various kinds of "matter" or
a cosmological
constant (inflation).
This is what we mean by mirage cosmology: the cosmological expansion is
not
due to energy density on our universe but somewhere else.
This can be either on other branes (that can be represented
qualitatively
by a black hole background) or in the bulk.

There is a different limit in which  our universe is moving to a  region
of
ultra-weak bulk fields,  in which case the
matter density on it alone drives the cosmological expansion, in the
traditional fashion.
We will show that this limit is equally well described in our setup.

The holographic principle and AdS/CFT
correspondence ideas, are providing novel ways to treat old mechanisms.
In the case of cosmological expansion, supersymmetry is very softly
broken
and it is expected that the probe
approximation may be valid even in the case where the source is not
hierarchically heavier than the probe \cite{2}.
It was observed that for a D-brane moving in the background of a
black D-brane the word-volume theory has an effective speed of light
which
is field dependent \cite{2}. Once the probe brane is in geodesic motion
the varying speed of light is equivalent to cosmological expansion on the
probe brane.

There are two possibilities to be explored in relation with the bulk
geometry.
The bulk may not be compact (but there is a mass gap \cite{gibb} or some
way it
makes low lying higher-dimensional gravitons   unobservable
\cite{susu}).
Then, there is no bound on the mass of D-branes.
If the bulk is compact there are charge neutrality constraints
that must be satisfied and they constrain the brane configurations.
In simple situations they limit the number of D-branes.
A typical example is the D9-branes in type-I string theory whose number
is
limited to 32.

Here we will mostly focus in the non-compact case.
In the case where the space is compact the generalization (and
limitation) of
our arguments will be straightforward.
For regions which are small compared to the size of the compact space
the
description of geodesics is accurate.
When a geodesic reaches distances comparable with the compact size we
must use
the form of bulk solution which is periodic (and can be constructed as
an
infinite periodic array of non-compact solutions).
Using such a matching formula the full geodesics can be studied.
This will imply that in such a context the most probable cosmological
evolution is a bouncing one.

Thus, the central idea is that the universe brane is moving into the bulk
background fields of other branes of the theory.
The motion of the brane follows thus, a classical geodesic
in the bulk geometry.
The prototype branes we are using here are Dp-branes with maximal
supersymmetry. Any realistic type-I ground-state can be viewed in big
region of
moduli space as intersecting such branes.
Moreover we know well the coupling of world-volume fields to the bulk
supergravity fields.

There are two steps in the procedure:
\begin{itemize}
\item
Determine the brane motion by solving the world-volume field equations
for the
scalar fields determining the position of the brane in the bulk

\item Determine the induced metric on the brane which now becomes an
implicit
function  of time. This gives a cosmological evolution in the induced
brane
metric. This cosmological evolution can be reinterpreted in terms of
cosmological ``mirage" energy densities on the brane via a Friedman-like
equation.
The induced metric on the brane is the natural metric felt by the observers
on the brane. We assume that our universe lives on the brane and is made off
open string fluctuations.

\item Similarly one can determine the other interactions on the brane

\end{itemize}

An important reminder here is that the cosmological evolution is
$not$ driven by four-dimensional gravity on the brane.
Our analysis indicates that potential inflationary models
where the position scalar and its potential is used to generate
inflation
on the brane via conventional four-dimensional gravity couplings might
not after all generate the sought-after inflation.
A necessary condition is that the minimum of the potential
(equal to the brane tension) to be above the BPS limit.

We have analyzed various combinations of branes and simple background
fields.
They correspond to a stack of Dp-branes on and out of extremality (black
Dp-branes) as well as with additional constant antisymmetric
tensor backgrounds. They provide a cosmological evolution on the probe
brane
that can be simulated by various types of mirage matter on the brane.
Most prominent is radiation-types ($w$=1/3) or massless scalars ($w$=1).
It should be stressed however that at small scale factor size, there are
many
exotic types of mirage matter including $w$ values that are outside the
range
$|w|\leq 1$ required by four-dimensional causality.
We interpret this as an indication that superluminal (from that
four-dimensional points of view) "shocks" are possible in such
cosmologies.
Superluminal signal propagation in a brane-world context
have been recently pursued independently in \cite{free,kal}.

   Another peculiarity is that individual densities of mirage dilute
matter can
be negative (without spoiling the overall positivity at late times).
   We think that this is linked to the fact that in this type of
cosmology
   the initial singularity is an artifact of the low energy description.

   This can be seen by studying brane motion in simple spaces like
$AdS_5\times
S^5$ which are globally non-singular.
   The induced cosmological evolution of a brane moving in such a space
has a
typical expansion profile due to radiation and an initial singularity
(from the
four-dimensional point of view).
However this singularity is an artifact.
At the point of the initial singularity the universe brane joins a
collection of
parallel similar branes and there is (non-abelian) symmetry enhancement.
The
effective field theory breaks down and this gives rise to the
singularity.

The next obvious question is how ``real" matter/energy densities on the
brane
affect its geodesic motion and consequently the induced cosmological
evolution.
This can be studied by turning on electromagnetic energy on the brane. We
find a solution of the moving brane with a covariantly constant electric field.
We do show that this gives an additional   effect  on the
cosmological evolution
similar to the analogous problem of radiation density in four-dimensions.
Although  an electric field is an unrealistic cosmological background
the solution we obtain is valid when the electric energy density is
thermal (and thus isotropic) in nature.
This indicates that the formalism we present is capable of handling the
most general situation possible, namely cosmological evolution driven by bulk
background fields (mirage matter) as well as world-volume energy
densities (real matter).

As it was first pointed out in \cite{2} this context allows
for an arbitrarily small residual cosmological constant on the universe brane.

The structure of the paper is as follows.
In section 2 we develop the formalism for D-brane geodesics in
non-trivial bulk
metric as well as RR form. The induced cosmological evolution is also
derived.
In section 3 we study the influence on the geodesics of an additional
constant NS antisymmetric tensor background.
In section 4 we derive the geodesic equations in the presence of an
electric field background on the probe brane.
In section 5 we study the Friedman-like equations for various examples
of bulk fields.
In section 6 we discuss the nature and fate of the initial
(cosmological)
singularity.
In section 7 we briefly discuss how the cosmological setup presented can be incorporated in string theory.
Finally section 8 contains our conclusions and open problems.

\section{Brane geodesics}
\setcounter{equation}{0}

In this section we will consider a probe D-brane moving in a generic
static,
spherically symmetric
background.
The brane will move in a geodesic. We assume the brane to be light
compared to
the background
so that we will neglect the back-reaction.
The idea here is that as the brane moves in a geodesic, the induced
world-volume metric
becomes a function of time, so that from the brane ``residents'' point
of view
they are
living in a changing (expanding or contracting) universe.

The simplest case corresponds to a D3-brane and we will focus mostly on
this case. Later we will consider Dp-branes with $p>3$ and $p-3$
coordinates compactified.
The metric of a D3-brane may be parameterized
as
\be
ds^2_{10}=g_{00}(r)dt^2+g(r)(d\vec x)^2+g_{rr}(r)dr^2+g_S(r)d\Omega_{5}
\label{mm}
\ee
and we may also generically have a
dilaton $\phi(r)$ as well as a RR background $C(r)=C_{0...3}(r)$ with a
self-dual field strength.
The probe brane will in general start moving in this background
along a geodesic and its dynamics is governed by the DBI action.
In the case of maximal supersymmetry it is given by
\be
S=T_3\int d^{4}\xi e^{-\phi}\sqrt{-det(\hat G_{\alpha\beta} +(2\pi
\alpha')F_{\alpha\beta}-B_{\alpha\beta})}+
T_3\int d^{4}\xi  ~\hat C_4 +{\rm
anomaly~~ terms} \label{action}
\ee
where we have ignored the world-volume fermions.
The emmbeded data are given by
\be
 \hat G_{\alpha\beta}=G_{\mu\nu}{\partial x^{\mu}\over
\partial\xi^{\alpha}}{\partial x^{\nu}\over
\partial\xi^{\beta}}\,
\ee
etc.
Due to reparametrization invariance, there is a gauge freedom which may
be
fixed by choosing the static gauge, $x^{\alpha}=\xi^{\alpha}$
$\alpha=0,1,2,3$.
A generic motion of the probe D3-brane will  have a non-trivial
angular
momentum in the transverse directions.
In the static gauge the relevant  (bosonic) part of the brane
Lagrangian reads
\be
L=\sqrt{g(r)^3[|g_{00}|-g_{rr}\dr^2-g_S(r)h_{ij}\dt^i\dt^j]}-C(r)
\label{the}\ee
where
$h_{ij}(\varphi)d\varphi^i\varphi^j$ is the line element of the unit
five-sphere.
For future purposes (generality) we will parametrize the Lagrangian as
\be
{\cal L}=\sqrt{A(r)-B(r)\dot r ^2-D(r)h_{ij}\dt^i \dt^j}-C(r)
\label{l}
\ee
with
\be
A(r)=g^3(r)|g_{00}(r)|e^{-2\phi}\sp B(r)=g^3(r)g_{rr}(r)e^{-2\phi}
\sp D(r)=g^3(r)g_S(r)e^{-2\phi}
\label{sub}\ee
and $C(r)$ is the RR background.
The problem is effectively one-dimensional and can be solved
by quadratures.The momenta are given by
\begin{eqnarray}
&& p_r=-{B(r) \dot r\over \sqrt{A(r)-B(r)\dot r ^2-D(r)h_{ij}\dt^i \dt^j}}\\ \nonumber &&
p_i==-{D(r) h_{ij}\dt^j\over
\sqrt{A(r)-B(r)\dot r ^2-D(r)h_{ij}\dt^i \dt^j}}
\end{eqnarray}
The angular momenta
as well as the  Hamiltonian
\be
H=-E=C-{A(r)\over
\sqrt{A(r)-B(r)\dot r ^2-D(r)h_{ij}\dt^i \dt^j}}
\label{the1}\ee
are conserved.
The conserved total angular momentum (SO(5) quadratic Casimir in our
case)
is $h^{ij}p_ip_j=\ell^2$ and
\be
h_{ij}\dt^i\dt^j={\ell^2(A(r)-B(r)\dot r^2)\over D(r)(D(r)+\ell^2)}
\ee
Thus, the final equation for the radial variable is
\be
\sqrt{{D\over D+\ell^2} (A(r)-B(r)\dr^2)}=
{A(r)\over E+C(r)}\, .
\ee
In summary,
\be
\dr^2={A\over B}\left(1-{A\over (C+E)^2}{D+\ell^2\over D}\right)\sp
h_{ij}\dt^i\dt^j={A^2\ell^2\over D^2(C+E)^2}
\label{solu}\ee

Here we see that we have the constraint that $C(r)+E\geq 0$ for the
allowed values of r.
A stronger condition can be obtained from (\ref{solu})
\be
{A\over B}\left(1-{A\over (C+E)^2}{D+\ell^2\over D}\right)\geq 0
\label{qu}\ee
Note that for slow motion (non-relativistic limit),
 we can expand the square root in (\ref{l}) to obtain
\be
{\cal L}_{n.r.}=\sqrt{A(r)}-C(r)-{1\over 2}{B(r)\over\sqrt{A(r)}} \dot r
^2
-{1\over 2}{D(r) \over \sqrt{A(r)}}h_{ij}\dt^i \dt^j\label{l1}
\ee
which is equivalent to a particle
moving in a potential of a central force.
The analogous solution is now
\be
\dr^2 =2\left[E+C-
\sqrt{A}\right]{\sqrt{A}\over B}-{A\ell^2\over BD}
\ee
The non-relativistic limit is valid when ${A\over (C+E)^2}\simeq 1$ and
$D(r)>>\ell^2$.

The induced four-dimensional metric on the 3-brane universe is
\be
d\hat s^2=(g_{00}+g_{rr}\dr^2+ g_Sh_{ij}\dt^i \dt^j)dt^2+g(d\vec x)^2
\label{met}
\ee
and upon substituting from (\ref{sub}) it becomes
\be
d\hat s^2=-{g^2_{00}g^3 e^{-2\phi}\over (C+E)^2}dt^2+g(d\vec x)^2
\ee
We can define the cosmic time $\eta$ as
\be
d\eta={|g_{00}|g^{3/2}e^{-\phi}\over |C+E|}dt
\ee
so that the universe metric is
\be
d\hat s^2=-d\eta^2+g(r(\eta))(d\vec
x)^2
\label{ff}\ee
The cosmic time is the same as the proper time of the universe brane.
Equation (\ref{ff}) is the standard form of a flat expanding universe.
We can now derive the analogues of the four-dimensional Friedman
equations
by defining the scale factor as
$a^2=g$.
Then,
\be
\left({\dot a\over a}\right)^2={(C+E)^2g_Se^{2\phi}-|g_{00}|(g_Sg^3+
\ell^2e^{2\phi})\over
4|g_{00}|g_{rr}g_Sg^3}
\left({g'\over g}\right)^2
\label{ffr1}\ee
where the dot stands for derivative with respect to cosmic time while
the
prime
stands for derivative with respect to $r$.
The right hand side of (\ref{ffr1}) can be interpreted in terms of an
effective
matter density on the probe brane
\be
{8\pi \over 3}\rho_{\rm eff}= {(C+E)^2g_Se^{2\phi}-|g_{00}|(g_Sg^3+
\ell^2e^{2\phi})\over
4|g_{00}|g_{rr}g_Sg^3}
\left({g'\over g}\right)^2
\label{den}
\ee
We have also
\be
{\ddot a\over a}=\left(1+{g\over g'}{\partial\over \partial r}\right)
{(C+E)^2g_Se^{2\phi}-|g_{00}|(g_Sg^3+
\ell^2e^{2\phi})\over
4|g_{00}|g_{rr}g_Sg^3}
\left({g'\over g}\right)^2=\left[1+{1\over 2}a{\partial\over \partial
a}\right]{8\pi \over 3}\rho_{\rm eff}
\ee
By setting the above equal to $-{4\pi \over 3}(\rho_{\rm eff} +3p_{\rm
eff})$
we can define the effective pressure $p_{\rm eff}$.

In terms of the above we can calculate the apparent scalar curvature of
the four-dimensional universe as
\be
R_{4-d}=8\pi(\rho_{\rm eff}-3p_{\rm eff})=
8\pi\left(4+a\partial_a\right)\rho_{\rm eff}
\label{curv}\ee

In the non-relativistic limit the induced metric is approximated by
\be
\left. d\hat s^2\right|_{n.r.}\simeq
g_{00}\left[1-2\left({(C+E)e^{\phi}\over
\sqrt{g^3|g_{00}|}}-1\right)-{\ell^2 e^{2\phi}/g^3
g_S}\right]dt^2+g(d\vec x)^2
\ee
while the effective density becomes
\be
\left. {8\pi\over 3}\rho_{\rm eff}\right|_{n.r}\simeq {1\over
4g_{rr}}\left({g'\over g}\right)^2\left[\left({(C+E)e^{\phi}\over
\sqrt{g^3|g_{00}|}}-1\right)-{\ell^2 e^{2\phi}\over g^3 g_S}\right]
\ee
As we will see in specific examples later on, this approximation is
usually valid for motion far away from the source corresponding to large
(late) time of the scale factor.
Moreover, going back towards what is the initial singularity,
relativistic corrections become increasingly important.

The discussion above may easily be generalized for
the geodesic motion of a probe Dp-brane in the field
of a Dp'-brane with $p'>p$. In this case, the Dp'-brane metric is of the
form
\be
ds^2_{10}=g_{00}(r)dt^2+g(r)(d\vec x_{p'})^2+g_{rr}(r)dr^2+g_S(r)
d\Omega_{8-p'}\, , \label{metricp'}
\ee
and there exist in general a non-trivial dilaton profile $\phi=\phi(r)$
as well as a RR $p'+1$ from $C_{p'+1}$. The Dp-brane probe in this
background
will
feel only gravitational and dilaton  forces since it has no $p'$-brane
charge.
Its motion will then determined by the DBI action
\be
S_p=T_p\int d^{p+1}\xi e^{-\phi}\sqrt{-det(\hat G_{\alpha\beta})}
\label{WZ}
\ee
In the static gauge and for a generic motion
with non-trivial angular-momentum in the transverse directions of the
Dp'-brane
we find that the Lagrangian turns out to be
\be
{\cal L}=\sqrt{A_p(r)-B_p(r)\dot r ^2-D_p(r)h_{ij}\dt^i \dt^j}
\label{l5}
\ee
where now
\be
A_p(r)=g^p(r)|g_{00}(r)|e^{-2\phi}\sp B_p(r)=g^p(r)g_{rr}(r)e^{-2\phi}
\sp D_p(r)=g^p(r)g_S(r)e^{-2\phi}\, .
\ee

Proceeding as before, we find
the induced metric on the Dp-brane,
\be
d\hat s^2=(g_{00}+g_{rr}\dr^2+g_Sh_{ij}\dt^i \dt^j)dt^2+g(d\vec x_p)^2
\label{met1}
\ee
and upon substitution becomes
\be
d\hat s^2=-{g^2_{00}g^pe^{-2\phi}\over E^2}dt^2+g(d\vec x_p)^2
\ee
We can define now the cosmic time $\eta$ as
\be
d\eta={|g_{00}|g^{p/2}e^{-\phi}\over |E|}dt
\ee
so that the universe metric is $d\hat s^2=-d\eta^2+g(r(\eta))(d\vec
x)^2$.
The analogues of the p+1-dimensional Friedman equations
are determined by defining the scale factor as
$a^2=g$.
As before we obtain a Friedman type equation with an effective density
given by
\be
{4\pi \over 3}\rho_{\rm eff}=
{E^2g_Se^{2\phi}-|g_{00}|(g_Sg^p+
\ell^2e^{2\phi})\over
4|g_{00}|g_{rr}g_Sg^p}
\left({g'\over g}\right)^2
\label{pp'}\ee

Note that in the case $p=p'$, there exist the additional WZ term
$T_p\int {\hat C}_{p+1}$ in the
action(\ref{WZ}) which modifies the equations of motion of the probe
brane
as well as the induced metric. This modification is nothing than the
shift
$E\to E+C$ where $C=C_{0...p}$.

\section{Branes with constant B-fields}
\setcounter{equation}{0}

Interesting dynamics may also arise by turning on the antisymmetric
NS/NS
field $B={1\over 2} B_{\mu\nu}dx^\mu\wedge dx^\nu$
on the macroscopic Dp-brane. In particular, a constant
$B$ will not affect the background since it enters in the field
equations via its field strength $H=dB$ which is zero for a constant
$B$.
However, a probe brane feels not $H$ but the antisymmetric field itself
through the coupling
\be
S=T_p\int d^{p+1}\xi e^{-\phi}\sqrt{-det(\hat G_{\alpha\beta} -
\hat B_{\alpha\beta})}+T_p\int d^{p+1}\xi  ~\hat C_{p+1} +{\rm
anomaly~~ terms}
\ee
where
\be
 \hat B_{\alpha\beta}=B_{\mu\nu}{\partial x^{\mu}\over
\partial\xi^{\alpha}}{\partial x^{\nu}\over
\partial\xi^{\beta}}\, .
\ee
We will assume again that there exist a macroscopic Dp'-brane with a
background metric as in eq.(\ref{metricp'}) and a probe
Dp-brane with $p<p'$.  In addition to the dilaton
$\phi(r)$ and RR form $C_{p'+1}$, the Dp'-brane now
 supports a constant NS/NS two-form
in its world-volume which we take to be
is $B=b dx^{p-1}\wedge dx^{p}$.
In general an antisymmetric tensor in the direction of the universe
brane will break isotropy.
However, in the case where the universe brane is higher-dimensional
with some of the directions compactified (which is the realistic
situation in type-I string theory) then we can turn on a component
$B_{0\theta}$ where $\theta$ is one of the compact directions. Such
an antisymmetric tensor does not break isotropy in the effective
 four-dimensional universe.

In the case at hand the induced
antisymmetric two-form on the probe Dp-brane and its motion is
determined by
the Lagrangian
\be
{\cal L}=\sqrt{K_p(r)-L_p(r)\dot r ^2-N_p(r)h_{ij}\dt^i \dt^j}
\label{lll}
\ee
where now
\begin{eqnarray}
&&K_p(r)=g(r)^{p-2}(g(r)^2+b^2)|g_{00}(r)|e^{-2\phi}\sp
L_p(r)=g(r)^{p-2}
(g(r)^2+b^2)
g_{rr}(r)e^{-2\phi}\, , \nonumber\\ &&
N_p(r)=g(r)^{p-2}(g(r)^2+b^2)g_S(r)e^{-2\phi}\, .
\end{eqnarray}
First integrals of the equations
of motion are now
\be
\dr^2={K_p\over L_p}\left(1-{K_p (N_p+\ell^2) \over N_pE^2}\right)\, ,
{}~~~ h_{ij}\dt^i \dt^j={K_p^2\ell^2\over N_p^2E^2}\, ,\label{quuu}
\ee
while the induced metric will still be given by eq.(\ref{met}).
Proceeding as
before we find that the Friedman-like equations are
\be
\left({\dot a\over a}\right)^2={E^2g_Se^{2\phi}-|g_{00}|\left(g_Sg^{p-2}
(g^2+b^2)+
\ell^2e^{2\phi}\right)\over
4|g_{00}|g_{rr}g_Sg^{p-2}(g^2+b^2)}
\left({g'\over g}\right)^2
\label{fr1}\ee
In addition we have
\be
{\ddot a\over a}=\left(1+{g'\over 4g}{\partial\over \partial r}\right)
{E^2g_Se^{2\phi}-|g_{00}|\left(g_Sg^{p-2}
(g^2+b^2)+
\ell^2e^{2\phi}\right)\over
4|g_{00}|g_{rr}g_Sg^{p-2}(g^2+b^2)}
\left({g'\over g}\right)^2
\ee
Consequently, the effective energy density on the probe brane turns out
to be
\be
{4\pi \over 3}\rho_{\rm eff}={E^2g_Se^{2\phi}-|g_{00}|\left(g_Sg^{p-2}
(g^2+b^2)+
\ell^2e^{2\phi}\right)\over
4|g_{00}|g_{rr}g_Sg^{p-2}(g^2+b^2)}
\left({g'\over g}\right)^2\, .
\ee

In the case $p=p'$, the probe brane feels also the RR form so that
the effective  energy density turns out to be
\be
{4\pi \over 3}\rho_{\rm
eff}={(C+E)^2g_Se^{2\phi}-|g_{00}|\left(g_Sg^{p-2}
(g^2+b^2)+
\ell^2e^{2\phi}\right)\over
4|g_{00}|g_{rr}g_Sg^{p-2}(g^2+b^2)}
\left({g'\over g}\right)^2\, .
\label{UFO}
\ee

\section{Electric Fields on the brane}
\setcounter{equation}{0}
\def\d{\partial}
\def\l{\lambda}
\def\dr{\dot r}
\def\E{{\cal E}}

Let us now assume that  we turn on an electric field on the probe
D3-brane.
This obviously breaks isotropy on the universe and thus this is not a realistic
configuration for cosmological purposes.
Our aim here is different. We would like to show that when there
energy density due to the  gauge fields of the brane, our approach
takes them appropriately into account and they will affect the
evolution of the cosmological factor.
Moreover, it is not difficult to argue that if the gauge-field energy density
is thermal in nature (and thus not isotropy breaking) this will not
affect our conclusions provided we substitute $\vec E^2\to <\vec E^2>$ etc.
The end result turns out to be  that our approach takes into account
 properly both mirage and real energy densities and one can eventually
 study the transition between the two.

When we keep track of the gauge fields,
the action for the D3-brane is given in
(\ref{action})
and for the background in (\ref{mm}) the Lagrangian takes the form
\begin{equation}
{\cal L}=\sqrt{A-B\dr^2-\E^2g^2} - C
\end{equation}
where $\E^2=2\pi\alpha'E_iE^i$ and $E_i=-\d_tA_i(t)$ in the
$A_0=0$ gauge and $A,B$ are given in (\ref{sub}).
For simplicity we focus on radial motion.

The equations of motions for the electric field turn out to be
\be
\d_t\left(g^2E_i\over \sqrt{A-B\dr^2-\E^2g^2}\right)=0.
\ee
and we find that
\be
E_i={\mu_i\over g}
\sqrt{A-B\dr^2\over \mu^2+g^2}\, , ~~~ \E^2={\mu^2\over
g^2}{A-B\dr^2\over
g^2+\mu^2}, \label{e}
\ee
where $\mu_i$ is an integration constant and
$\mu^2=(2\pi\alpha')\mu_i\mu^i$.
In the case $\dr=0$, $E_i$ is constant as it is required by
ordinary Maxwell
equations.
A first integral is given by
\be
\dr^2={A\over B} \left(1-{A(1+\mu^2g^{-2})\over (C+E)^2}\right).
\label{r}
\ee
Using (\ref{r}), we obtain
\be
\E^2={\mu^2\over g^4}{A^2\over
(C+E)^2}.
\label{ee}
\ee
The induced metric on the probe D3-brane turns out to be
\be
d\hat s^2=-{g_{00}^2 g^5e^{-2\phi}\over (C+E)^2(\mu^2+g^2)}dt^2+g(d\vec
x)^2
\ee
and by defining the cosmic time $\eta$ along similar lines
we obtain the following  analog of the
Friedman
equations
\be
\left({\dot a\over
a}\right)^2={{g^2(C+E)^2\over (g^2+\mu^2)}-|g_{00}|g^3e^{-2\phi}\over
4|g_{00}|g_{rr}g^3 e^{-2\phi}}
\left({g'\over g}\right)^2
\label{fr13}
\ee

We note that the dominant contribution to $(\dot a/ a)^2$ from the
electric
field is of order $\E^2$ as can be seen from eq.(\ref{ee}) and
thus
proportional to the energy density $\rho \sim \E^2$. It should
also be noted
that there exist a limiting value for the electric field
\cite{FTs},\cite{BP}
which is
\be
\E^2\leq A^2g^{-2}
\ee

The gauge invariance of the bulk antisymmetric tensor is closely tied
with
that of the world-volume gauge fields.
This is in agreement with the observation that, comparing equations
(\ref{UFO})
(with $\ell=0$)
and (\ref{fr1}) that an electric field on the brane or a constant
antisymmetric
tensor in bulk produce similar cosmological evolution.

\section{Cosmology of a probe D3-brane in various backgrounds}
\setcounter{equation}{0}
In this section we will analyse several concrete bulk configurations
and elaborate on the induced cosmological expansion on the universe brane.

\noindent
{\bf AdS$_5$ black hole}
\vspace{.2cm}

The near-horizon geometry of a macroscopic D3-brane is $AdS_5\times
S^5$.
Once the brane is black we obtain the associated black hole solution
with metric
\be
ds^2={r^2\over L^2}\left(-f(r) ~dt^2+(d\vec x)^2\right)+{L^2\over
r^2}{dr^2\over f(r)}+
L^2d\Omega_5^2\, , \label{met3}
\ee
$f(r)=1-\left({r_0\over r}\right)^4$
and RR field $C=C_{0...3}=\left[{r^4\over L^4}-{r_0^4\over
2L^4}\right]$.
The constant part can be eventually absorbed into a redefinition of the
parameter E.

By using eqs.(\ref{mm},\ref{den}), the
effective density on a probe D3-brane in the above background turns out
to be
 \be
{8\pi \over 3}\rho_{\rm eff}={1\over L^2}\left[ \left(1+{E\over
a^4}\right)^2-\left(1-\left({r_0\over L}\right)^4{1\over
a^4}\right)\left(1+{\ell^2\over L^2}{1\over a^6}\right)\right]
\label{met4}\ee
When the brane is falling towards the black-brane the universe is
contacting
while if it moving outwards, it is expanding.
Far from the black-brane $\rho_{eff}\sim a^{-4}$. In this regime the
brane
motion
produces a cosmological expansion indistinguishable with the one due to
(dilute)
radiation on the brane.
As one goes backward in time there is a negative energy density $\sim
a^{-6}$
 controlled by the angular momentum $\ell$.
It corresponds to $\rho_{\rm eff}=p_{\rm eff}$, relation characteristic
of  a massless scalar.
Although the density is negative the overall effective density remains
non-negative
for $0<a<\infty$.
At earlier times the factor $\sim a^{-8}$ dominates corresponding to
dilute matter with $p=w~\rho$ and $w=5/3>1$. Such a behavior is
unattainable
 by real matter on the brane since causality implies that $|w|\leq 1$.
Finally, at very early times the evolution is dominated
by mirage density with $w=7/3$.

More generally we can consider the D3-brane moving in the background of
a $p>3$
black brane.
This problem has been considered in a previous section.
In this case we obtain an effective density using (\ref{pp'}) with p=3:
\be
{8\pi\over 3}\rho_{\rm eff}=\left({7-p\over 4L}\right)^2~a^{2(3-p)\over
7-p}~
\left[{E^2\over a^{2(7-p)}}
-\left(1-\left({r_0\over L}\right)^{7-p}{1\over a^4}\right)
\left(1+{\ell^2\over L^2}{1\over a^{2(5-p)+{8\over 7-p}}}\right)\right]
\label{pb}\ee
As is obvious from the above, that the universe brane cannot go far away
from
the black-brane. It bounces back at some finite value of the scale
factor.
, this described a closed universe where the deceleration is provides
by
bulk fields
rather than curvature on the brane.
Particularly, interesting is the case $p=5$ where (\ref{pb}) reads
\be
 {8\pi\over 3}\rho_{\rm eff}={1\over 4L^2}~{1\over a^{2}}~
\left[{E^2\over a^{4}}
-\left(1-\left({r_0\over L}\right)^4{1\over a^4}\right)
\left(1+{\ell^2\over L^2}{1\over a^{4}}\right)\right]
\ee
The term $-a^{-2}$ produces an effect similar to positive curvature and
slows
the expansion.
The terms $a^{-6}$ simulate the density of a massless scalar $w=1$.
This density increases with $E$ and $r_0$ while the angular momentum
tends to
decrease
the density.
Moreover there is also a component with $w=7/3$
In the case $\ell=r_0 =0$, the universe expands until it eventually
stops and
eventually recollapses.
The general picture with recollapse is  also true for all $p>3$.
For $p=4$ we obtain the effective indices $w=-7/9,5/9,11/9,7/9,19/9$.
For $p=6$ we obtain $w=1,5/3,7/3,3,13/3$.

\vspace{.2cm}

\noindent
{\bf Dp-black brane}
\vspace{.2cm}

The metric is
\be
ds^2=H_p^{-1/2}\left(-fdt^2+(d\vec x)^2\right)+H_p^{1/2}\left({dr^2\over
f}+
r^2d\Omega_{8-p}^2\right)\, ,
\ee
with $H_p=1+{L^{7-p}\over r^{7-p}}$, $f=1-{r_0^{7-p}\over r^{7-p}}$.
The RR field is $C=\xi{1-H_p\over H_p}$ with
$\xi= \sqrt{1+{r_0^{7-p}\over L^{7-p}}}$
and the dilaton is $e^{\phi}=H_p^{(3-p)/4}$. A probe D3-brane in this
background
has a cosmological evolution driven by the effective density
\be
{8\pi \over 3}\rho_{\rm eff}={(1-a^4)^{5/2}\over L^2}\left[{(E+\xi
a^4)^2\over
a^8}
-\left(\xi^2-{\xi^2-1\over a^4}\right)\left(1+{\ell^2\over
L^2}{\sqrt{1-a^4}\over a^6}\right)\right]
\ee
when $p=3$ and
\be
{8\pi \over 3}\rho_{\rm eff}={(7-p)^2\over 16L^2}
a^{2(3-p)\over (7-p)}(1-a^4)^{2(8-p)\over (7-p)}\left[{E^2\over
a^{2(7-p)}}
-\left(\xi^2-{\xi^2-1\over a^4}\right)\left(1+{\ell^2\over
L^2}{(1-a^4)^{2\over
(7-p)}
\over a^{{2(5-p)+{8\over 7-p}}}}\right)\right]
\ee
when $p>3$

Here there is a limiting size for the scale factor namely $0\leq a\leq
1$.
the maximum value is in general a free parameter that can always be
scaled to
one in the equations.
For $a<<1$ the cosmological evolution is similar to the one discussed
for
the
near-horizon geometries.
When $a\simeq 1$ there is a different type of evolution.
Denoting $a^4=1-\epsilon$, $\epsilon<<1$
we obtain (for $p<7$)
\be
\dot \epsilon ={|7-p|\over L}\sqrt{E^2-1}~\epsilon^{8-p\over 7-p}\sp
\Rightarrow\sp
\epsilon={L^{7-p}\over (\sqrt{E^2-1})^{7-p}}t^{-(7-p)}
\ee
This is the typical behavior of asymptotically flat solutions.
The case $p=7,8$ are not asymptotically flat.
We will discuss $p=8$ as an example  below.
The effective density is in this case
\be
{8\pi \over 3}\rho_{\rm eff}={1\over 16L^2}
a^{10}\left[E^2~a^2
-\left(\xi^2-{\xi^2-1\over a^4}\right)\left(1+{\ell^2\over L^2}{a^{14}
\over (1-a^4)^{2}}\right)\right]
\ee
For small $a$ the dominant term in the evolution is $a^6$.
For later times the scale factor will bounce depending on parameters
before
$a=1$.

As a conclusion for the asymptotic geometries of spherically symmetric
branes,
the evolution equation for the scale factor is different from standard
evolution due
to some kind of density on the brane.
On the other it will be expected that once the universe brane is far
away from
the source and the gravitational and other fields are weak,
in this regime the dominant source of cosmological expansion will
be the matter density on the brane.

\bigskip
\noindent
{\bf Near horizon of a D3-brane with a constant B-field}
\vspace{.2cm}

The near-horizon geometry of a D3-brane with a constant B-field is still
given
 by eq.(\ref{met3}).
By using eq.(\ref{UFO}) for $p=3$, we find that the
effective density on a probe D3-brane in the D3-brane  background
with a constant B-field turns out to be
 \be
{4\pi \over 3}\rho_{\rm eff}={1\over L^2}\left[\left(1+{E\over
a^4}\right)^2\left(1+{b^2\over a^4}\right)^{-1}-\left(1-{r_0^4\over
L^4}{1\over
a^4}\right)\left(1+{\ell^2\over L^2} {\left(1+{b^2\over
a^4}\right)^{-1}\over
a^6}\right)\right]
\ee

\bigskip
\noindent
{\bf Near horizon of a D3-brane with a world-volume electric field}
\vspace{.2cm}

It is not difficult to see that the effect of the electric field
affects the cosmological evolution as it would in a four dimensional
universe.

In the case of the near-extremal  D3-brane (\ref{met3}) we obtain
instead of (\ref{met4})
\be
{8\pi \over 3}\rho_{\rm eff}={1\over L^2}\left[ {\left(1+{E\over
a^4}\right)^2\over \left(1+{\mu^2\over a^4}\right)}-\left(1-\left({r_0\over
L}\right)^4{1\over a^4}\right)\left(1+{\ell^2\over L^2}{1\over
a^6}\right)\right]
\label{met5}\ee
We obtain the same evolution as with a constant bulk NS antisymmetric tensor.
This was expected on general grounds.
The extra electric field adds at late times an extra effective
density
$\Delta \rho=-{\mu^2\over a^4}$. 
The negative sign is relative to the overall energy that 
we had set to be negative, namely -E. 

This is a behavior that is expected: extra matter density on the brane,
affects
the geodesics (motion) , and this in turn affects the effective
expansion of
the brane universe.
Moreover, this gives for small fields (or large scale factors) effects
that are
similar to those  of a constant bulk NS
antisymmetric tensor.
Both produce an effect that can be interpreted as radiation in our
brane
universe.

\section{The resolution of the initial singularity}
\setcounter{equation}{0}

In four dimensions standard cosmological models always carry an initial
singularity. This is the point in the
past of the evolution where the scale factor $a(t)$ goes to zero so
that
all space-like sections of space-time collapse to a point.
This is a general feature and  powerful theorems have  established the
occurrence of the initial singularity for matter obeying the
energy-conditions \cite{HE}.
Since the latter are satisfied for all known forms of matter, the initial
singularity seems to be unavoidable.
The basic assumption in the above is that the full description is given
in terms of four-dimensional general relativity.
If one views general relativity as an effective theory of a more
fundamental theory, then the presence of the initial singularity may be
resolved in the fundamental theory.
One could argue that the singularity appears not because the fundamental
solution is singular but because the effective field theory used to
describe, is not valid in this regime \cite{kk,ni}. In particular, as discussed
for example in \cite{ni}, due to T-duality an initial singularity could
really correspond to a decompactification limit for the relevant low
energy (dual) modes.

In our context,  there are cases where ``mirage" energy violates
the standard energy conditions.
For example for dilute matter we have $p=w~\rho$ with $|w|\leq 1$ for causality.
In our the previous section we have found the mirage matter had
$|w|>1$ in most of the cases,
(as well as some components of the density being negative) leading
to the possibility of singularity-free evolution.

This can be seen in the simple example of the geodesic motion of a
probe D3-brane in the near-horizon of macroscopic D5-branes with metric
\be
ds^2={r\over L} \left(-dt^2+ d{\vec x}_5^2\right)+{L\over r}\left(dr^2+r^2
d\Omega_{3}\right)\, .
\ee
In this case, solving for $r(t)$ with $\ell=0$ we get
\be
r(t)=\frac{L\, |E|}{\cosh(t/L)}
\ee
The induced metric (\ref{met}) turns out to be
\be
ds= -\frac{|E|}{\cosh(t/L)^3}dt^2+\frac{|E|}{\cosh(t/L)}d\vec{x}_3^2
\ee

This  four-dimensional metric has a singularity for $t=\infty$ (initial singularity).
However, the higher dimensional geometry is regular.
What becomes singular is the embedding of the brane in the bulk.
Alternatively speaking, the four-dimensional singularity is smoothed out once the solution is lifted to higher dimensions.
This is a well known method for de-singularizing four-dimensional manifolds
and appears here naturally.

Similarly, we may consider the motion of a probe D3-brane in $AdS_5
\times S^5$ background.
Indeed we expect that for motion in this smooth manifold no real
singularity on the brane can be encountered.
Here we find from (\ref{qu}) and
for $\ell=0$  that $0\leq r^4<\infty$ for $E>0$. The proper
time $\eta$ is then given
\be
\eta={1\over 2L}\sqrt{2r^4+L^4}-{L\over 2} \, ,
\ee
defined such that $\eta=0$ for $r=0$ and thus $0\leq
\eta<\infty$. On the other hand, solving for the scale factor $a(\eta)$
we
find that
\be
\alpha(\eta)^4={8E\over L^2}\left((\eta+\eta_0)^2-{L^2\over 16}\right)\, ,
\ee
where $\eta_0$ is an integration constant. At $\eta=0$ we get that
\be
\alpha(0)^4={8E\over L^2}\left(\eta_0^2-{L^2\over 16}\right)\, ,
\ee
so that we obtain the standard singularity for $|\eta_0|>L/4$.
This is the point that the brane reaches $r=0$ which is a coordinate
singularity and otherwise a regular point of the $AdS_5$ space.
Again, the embedding becomes singular there.

 From the string theory picture we do understand that the initial
singularity
here corresponds to the probe brane coalescing with the other branes
that generate the bulk background. The effective field theory on the
brane is singular at this point because one has to take into account the
non-abelian modes
(zero length strings) that become massless.
The interpretation of the real initial singularity here is as a
breakdown of the effective low energy field theory description.

\section{Incorporation into string theory}

Orbifold or orientifold backgrounds in string theory have a brane interpretation.

A fixed hyperplane of an orientifold transformation can be thought of as
a bound state of an orientifold plane and a D-brane.
The orientifold plane carries no degrees of freedom localized on it, but
only some CP-odd couplings to cancel similar couplings of the bound
D-brane.
When moduli are varied, the D-brane(s) can move away from orientifold
planes.
In open string vacua, extra D-branes can participate in the structure of
the vacuum.

In fact, a similar interpretation can be given to ordinary orbifold
vacua of the heterotic and type II strings
\footnote{This is implicit in \cite{kut}.}.
The fixed planes of the orbifold are
bound states of an orbifold plane and an NS-brane for a $Z_2$ twist.
For a $Z_N$ twist there are $N-1$ non-coincident NS5-branes.
The degrees of freedom localized on the orbifold plane are essentially
composed of the twisted sector fields of the orbifold.
Turning on twisted moduli expectation values corresponds to moving the
NS5-branes away from the orbifold plane.

Thus, a typical (orbifold) vacuum of type I theory can be thought
of as a
collection of flat (toroidal) intersecting D-branes in a
ten-dimensional
 flat bulk space. Constraints have to be satisfied, notably tadpole
cancellation which ensures anomaly cancellation and reflects charge
neutrality in a compact space.
In the presence of some unbroken space-time supersymmetry, the
configuration
of D-branes is stable.
Stability is a non-trivial requirement in the case of broken
supersymmetry.
It is however known that supersymmetric (BPS) D-branes have velocity
dependent interactions \cite{bachas}.
Thus, there is non-trivial dynamics in the case that a D-brane is
disturbed away from a stable configuration.

In this context, the Standard Model gauge bosons as well as charged
matter
arise as fluctuations of the D-branes. We can thus consider the universe
(standard model) to be living on a collection of coincident branes,
while
hidden gauge interactions can be localized on other branes and gravity
as
well as other universal interactions is living in the bulk space.

The approximation which we are using is very useful in order to treat the
dynamics of the universe brane. This is the probe limit in which the
influence
of the probe brane source to the bulk fields is negligible.
This has been a natural and useful tool  \cite{prob}-\cite{2}
in order to understand issues in
the context of AdS/CFT correspondence \cite{mald}.
However, if we can approximate the collection of other branes
(except the universe brane) that form the string vacuum as a spherically
symmetric configuration, then our treatment is (classically) exact. The reason is
the  back-reaction to the bulk metric due to the probe brane modifies
the bulk metric for radii larger that the position of the brane.

\section{Conclusions and further directions}

We have pointed out that if our universe is a D-brane embedded in a
higher
dimensional bulk, the motion of the brane in nontrivial bulk
backgrounds
with a certain symmetry (spherical in our case) produces a homogeneous
and isotropic cosmological evolution on the universe brane.
By considering various bulk background fields, we have derived
Friedman-like equations. These provide cosmological evolution that can be attributed to
matter density on the universe brane, although they are due to motion in
nontrivial bulk fields. From the universe brane perspective such energy
density is a mirage. The only way to tell whether the energy density
driving
the evolution of the universe is in the universe itself is by direct
observation with conventional means (light for example).

It is also important to point out that we consider a situation where
the spatial sections of the universe brane are flat.
Unlike the typical four-dimensional case though, the effects of non-trivial
spatial curvature (a $a^{-2}$ term in the Friedman equation)
can be simulated by the brane motion.
We have seen in section 5 that a D3-brane moving the field of parallel
D5-branes produces an effect similar to having a negative constant curvature
in the spatial slice.

We should also point out here that it is also possible to consider situations
where the induced brane metric really does have positive constant curvature.
That should be produced by a bulk $B_{\mu\nu}$ field with $H_{\mu\nu\rho}
\sim \epsilon_{\mu\nu\rho}$. Finding the precise solution in IIA,B supergravity
is an interesting problem. such a  solution though is known in gauged-supergravity in five-dimensions.
The only difference in our equations is to replace the spatial section metric
 on the brane by a constant curvature one, and add the usual curvature contribution $-k/a^2$ to our Friedman-like equations.

When we have extra world-volume fields excited on the brane, these
affect the universe brane motion, the induced metric and thus the cosmological
evolution.
We have analysed the case of electromagnetic energy (electric fields)
and shown that it affects the cosmological evolution as normal radiation
would.
This indicates that the present formalism is capable of describing the
most general cosmological evolution on a universe brane.
This is driven by bulk fields (that we could label ``bulk energy") as
well as world-volume fields (``brane-energy").
The cosmological evolution in this context has the simple and appealing
interpretation of brane motion in the bulk.

Generically, the Friedman-like equations we find, contain components
that can be interpreted as dilute matter with $|w|>1$ that would otherwise
clash with causality in four-dimensions. This may be an indication that
the bulk can support superluminal signal propagation from the brane point of view.
Moreover, sometimes, some effective density coefficients can be negative
(without spoiling the positivity of the overall effective density).
This violates typical four-dimensional  positive energy conditions.

A Friedman type cosmological evolution is equivalent to a variable
speed of light in the context of non-dynamical gravity.
Cosmological implications of a variable light speed have been
recently discussed in \cite{slight}.
Our description here can be cast in that language \cite{2}.
In \cite{2} it was pointed out that a natural way of inducing a field-dependent
light velocity of brane is in the presence of a (cosmological) black brane.
Subsequent motion of the probe brane, either falling into the black brane,
or escaping outwards (a Hawking radiated brane ?) would make the velocity
of light field dependent and thus induce cosmological evolution.

The issue of the initial singularity has a natural resolution in this context.
Although the universe brane has a singular geometry at zero scale factor,
 this  is due to a singular embedding in the otherwise regular bulk space.
This resolution of singularities in higher dimensions is not new
but occurs naturally here.
Relativistic corrections are important when one approaches the initial
singularity. They  invalidate for example naive treatments of brane
fields when the universe brane approaches  other branes.

There are some interesting open problems in this line of thought.

\begin{itemize}

\item The first concerns the possibility of inducing and ending inflation
as a consequence of brane motion. This is similar in spirit with \cite{dvty}
but the context is different: here inflation is not generated by four-dimensional gravity on the brane.
This will provide for a natural source of inflation since the inflaton is
supplied automatically once we decide that our four-dimensional universe
is a soliton-like object in a higher-dimensional theory.

\item A second direction is finding bulk configurations that induce
a cosmological evolution on the brane similar to that of dust (w=0).
Such mirage matter might be a component of dark matter
on our universe. It is conceivable that  such ``mirage" matter could gravitate and trace
the visible (world-volume) matter as observation suggests.

\item Finding brane configurations in ten-dimensional supergravity
with non-zero spatial curvature is also an interesting problem.

\end{itemize}

\vskip 2cm

\centerline{\bf\large Acknowledgements}
\vskip 1cm

One of the authors (E.K.) would like to acknowledge the
hospitality and support
of the ITP in S. Barbara where part of this work was done.
He also thanks the organizers of the September 1999, TMR meeting in ENS,
Paris for their kind invitation to present these results.
We would like to thank I. Antoniadis, C. Bachas, G. Gibbons,
C. Kounnas, J. Lykken, E. Poppitz, J. Russo and T. Tomaras,
for discussions and suggestions.
This work was partially supported through a TMR contract
ERBFMRX-CT96-0090 of the European Union, a  $\Gamma.\Gamma .$E.T. grant, No.
97$E\Lambda$/71 and by the National Science Foundation
under Grant No. PHY94-07194.
We would also like to thank the referee for his remarks that helped clarify
some of the presentation in this paper.

\vskip 2cm

\centerline{\bf\large Note Added}
\vskip 1cm

While this work was written up, reference \cite{kra} appeared where
a similar problem is tackled with a different technique (namely using
the Israel matching conditions).
The resulting equ. (17) for brane motion in an AdS black-hole
(with $\sigma=\sigma_c, k=0$) matches our equation (5.2) with $\ell=0$ with
the identifications $E={\mu_+-\mu_-}/2L^2$, $(r_0/L)^4=2\mu_-/l^2$, $L=l$.
This is as expected since the energy of motion $E$ creates a discontinuity
of the black-hole horizon position $\mu$ in the bulk.
Here we have a direct D-brane description which requires $\sigma=\sigma_c$.

We also became aware of the work in ref. \cite{cr} which partly overlaps
with our work. There, the Israel matching conditions are also used to
 study brane motion and some interesting exact solutions are found, exhibiting cosmological evolution.

\end{document}